\documentclass[twocolumn,showpacs,amsfonts]{revtex4-1}
\usepackage{amsmath}
\usepackage{latexsym}
\usepackage{mathrsfs}
\usepackage{float}
\usepackage{amssymb}
\usepackage{amsfonts}
\usepackage{graphicx}
\usepackage{textcomp}
\usepackage{hyperref}
\usepackage{mathrsfs}
\usepackage{pifont}
\usepackage{color}

 1

\newcommand{\dummy}

\begin{document}
\title{Simulation of Transport around the Coexistence 
Region of a Binary Fluid}
\author{Sutapa Roy and Subir K. Das$^*$}
\affiliation{Theoretical Sciences Unit, Jawaharlal Nehru 
Centre for Advanced Scientific Research, Jakkur P.O, 
Bangalore 560064, India}
\date{\today}

\begin{abstract}
We use Monte Carlo and molecular dynamics simulations to 
study phase behavior and transport properties in a 
symmetric binary fluid where particles interact via 
Lennard-Jones potential. Our results for the critical 
behavior of collective transport properties, with 
particular emphasis on bulk viscosity, is understood via 
appropriate application of finite-size scaling technique. 
It appears that the critical enhancements in these 
quantities are visible far above the critical point. 
This result is consistent with an earlier report from 
computer simulations where, however, the authors 
do not quantify the critical singularity.
\end{abstract}
\maketitle

\section{\label{intro}Introduction}
\par
\hspace{0.2cm}Understanding of various thermodynamic 
and transport properties in fluids, particularly at 
phase transitions, is of fundamental importance 
\cite{hohenberg,ferrell,olchowy,luettmer,onuki1,
anisimov,onuki2,kadanoff,mistura,folk,burstyn1,burstyn2} 
and gained momentum recently \cite{anisimov,onuki2,hao,
jkb1,jkb2,sengers1,perez,fisher1,luijten,kim,ashton,
das1,gillis1,gillis2,das2,meier,salin,dyer,das3,das4,
chen,gross,mutoru,fang,kumar}. Compared to thermodynamic 
properties, issues related to transport are inadequately 
addressed, particularly for quantities associated with 
the collective behavior of the systems. Knowledge of 
transport properties is crucial to the complete 
understanding of phase transition. Clearly, the kinetics 
of phase transition \cite{onuki2,bray} is dictated by 
various transport coefficients. E.g., 
amplitude of growth as well as crossover from one growth 
mechanism to the other, in fluid phase separation, are 
decided by various transport coefficients \cite{bray,furukawa}. 
Not only for such nonequilibrium processes, understanding 
of transport properties is extremely important in 
equilibrium context as well.  In this paper, we study 
transport in a binary fluid, with particular focus on 
critical phenomena, via computer simulations.

\par
\hspace{0.2cm}Behavior of various thermodynamic properties 
are reasonably well understood in the vicinity of critical 
point \cite{zinn}. In studies of such static critical 
phenomena, computer simulations played an important role 
\cite{landau} despite the well known fact that one faces 
difficulty due to finite size of the systems. Essentially, 
in this approach one cannot pick up the correct value of 
the characteristic length scale that diverges at criticality. 
It has, however, been possible to overcome this difficulty 
via finite-size scaling analysis \cite{fisher2}. Understanding 
of simulation results, primarily Monte Carlo \cite{landau}, 
via this method, in addition to confirming predictions of 
renormalization group theory \cite{zinn}, high temperature 
expansion \cite{pelissetto}, etc., provided much further 
information.

\par
\hspace{0.2cm}As already stated in more general context, 
much less volume of the literature is devoted to the 
understanding of dynamic critical phenomena, compared to its 
static counterpart.  This is more true for computer simulations 
because of certain difficulties. Nevertheless, 
accurate theoretical predictions \cite{onuki1,anisimov,onuki2,hao,jkb1,
jkb2,kadanoff,mistura,folk} and state-of-the-art experiments 
\cite{gillis1,gillis2,burstyn1,burstyn2} do exist. As opposed 
to the thermodynamic properties, where finite-size effect is 
the only difficulty in computer simulations, another robust 
hurdle in dynamics is the critical slowing down. Computational 
studies in this area are rare due to this combined difficulty.

\par
\hspace{0.2cm}In this work we aim to study dynamics in a binary 
($A+B$) fluid in space dimension ($d$) three. Below we briefly 
review some basic aspects of critical phenomena, this being the 
primary focus of the paper, for both static and dynamic properties. 
It is well known that in the vicinity of a critical point various 
thermodynamic and transport properties show power-law singularities 
as a function of deviation of temperature ($T$) from the critical 
value ($T_{_c}$). Typical examples are susceptibility ($\chi$), 
correlation length ($\xi$), order-parameter ($m$), specific heat 
($C$), shear viscosity ($\eta$), bulk viscosity ($\zeta$), mutual 
diffusivity ($D_{_{AB}}$), etc. In terms of 
$\epsilon[=|T-T_{_c}|/T_{_c}]$, they are described as \cite{onuki1}
\begin {eqnarray}\label{intro1}
m\sim \epsilon^\beta,~\chi\sim\epsilon^{-\gamma},~C\sim\epsilon^{-\alpha},
~\xi\sim\epsilon^{-\nu},
\end {eqnarray}
\begin {eqnarray}\label{intro2}
~\eta\sim\xi^{x_{_\eta}},~\zeta\sim\xi^{x_{_\zeta}},
~D_{_{AB}}\sim \xi^{-x_{_D}}.
\end {eqnarray}
\par
\hspace{0.2cm}It has been established that for static 
critical phenomena exponents have unique values for 
paramagnetic to ferromagnetic transition, immiscibility 
transition in a binary solid or fluid, vapor-liquid 
transition, etc. and belong to the Ising universality 
class with \cite{zinn}
\begin {eqnarray}\label{exponent1}
\beta\simeq0.325,~\gamma\simeq1.239,~\alpha\simeq0.11,
~\nu\simeq0.63.
\end {eqnarray}
On the other hand, universality of dynamic critical 
phenomena is less robust. Nevertheless, it is expected 
that liquid-liquid and vapor-liquid transitions will fall 
in the same class, described by model-$H$, having exponent 
values \cite{onuki2}
\begin {eqnarray}\label{exponent2}
x_{_\eta}=0.068,~x_{_\zeta}=2.89,~x_{_D}=1.068,
\end {eqnarray}
predicted by dynamic renormalization group and mode 
coupling theories. 

\par
\hspace{0.2cm}Analogous to the relations combining the 
exponents in static phenomena, scaling relations, such as 
\cite{onuki2}
\begin {eqnarray}\label{exponent3}
x_{_D}=1+x_{_\eta},
\end {eqnarray}
exist in dynamics also. Eq. (\ref{exponent3}) can be 
verified from the generalized Stokes-Einstein-Sutherland 
relation \cite{onuki2, squires, hansen}
\begin {eqnarray}\label{exponent4}
D_{_{AB}}=\frac{R_{_D}k_{_B}T}{6\pi\eta\xi},
\end {eqnarray}
where $k_B$ is the Boltzmann constant and $R_D$ is another 
universal number \cite{luettmer} $\simeq 1.03$. Some more 
such relations involving static and dynamic exponents 
are \cite{onuki2}
\begin {eqnarray}\label{exponent5}
x_{_\zeta}=z-\frac{\alpha}{\nu},~z=d+x_{_\eta},
\end {eqnarray}
etc. In Eq. (\ref{exponent5}), $z$ is an exponent related 
to the divergence of the relaxation time ($\tau$) of the 
system at criticality as \cite{onuki1,landau}
\begin {eqnarray}\label{exponent6}
\tau \sim \xi^z\sim L^z,
\end {eqnarray}
where $L$ is the linear dimension of the system under 
consideration. A basic fact used in writing Eq. (\ref{exponent5}) 
is that at criticality $\xi$ is of the size of the system, 
so they scale with each other. This fact we will use in the 
finite-size scaling analysis later.

\par
\hspace{0.2cm}With increasing computer power, combined with 
appropriate application of finite-size scaling theory, it 
appears now that it is possible to simulate dynamic critical 
phenomena \cite{das3,das4}. In this paper, in addition to 
confirming the values of some of the dynamic exponents, we 
present molecular dynamics (MD) simulation results for the 
transport properties in a wide range of the parameter space 
around the two-phase coexistence curve. A particular focus 
in this work is on the understanding of critical behavior of 
bulk viscosity. An interesting finding is the wide temperature 
range over which we observe critical enhancement in the 
transport properties which was presented in a letter \cite{roy1}. 
However, without results in the close vicinity of the critical 
point, this claim about ``wide critical range'', to a good degree, 
is not justified. Even though computationally very challenging, we 
have achieved this objective and will present the results in this 
longer communication.

\par
\hspace{0.2cm}Rest of the paper is organized as follows. We 
describe the model and methodologies in Section II. Section 
III contains all our results. Finally, we conclude the paper in 
Section IV with a brief summary and discussion on future 
possibilities.

\section{Model and Methods}\label{model}
\par
\hspace{0.2cm}Following Refs. \cite{das3,das4}, we describe 
the model below. Here particles of equal mass ($m_p$) and 
diameter ($\sigma$) interact via the standard Lennard-Jones 
(LJ) pair potential 
\begin {eqnarray}\label{LJ1}
u(r)=4\varepsilon_{_{ij}}\Big[\Big(\frac{\sigma}{r}\Big)^{12}
-\Big(\frac{\sigma}{r}\Big)^{6}\Big],
\end{eqnarray}
where $r$ is the scalar distance between $i$th and $j$th 
particles between which the strength of interaction is 
$\varepsilon_{_{ij}}$. To reduce the cost of computation, we 
used the modified potential
\begin {eqnarray}\label{LJ2}
U(r)=u(r)-u(r_{_c})-(r-r_{_c})
{\frac {du}{dr}}\Big|_{{r}=r_{_c}}.
\end{eqnarray}
The cut-off distance $r_{_c}$ was set to $2.5\sigma$. The 
shifting of the potential to zero at $r_{_c}$ makes the 
potential continuous, however, leaves the force discontinuous 
at $r=r_{_c}$. The last term in Eq. (\ref{LJ2}) was introduced 
to avoid this problem. Here we mention that such modification 
of the potential will certainly affect the non-universal 
quantities, e.g., $T_{_c}$, but will keep the critical 
universality unaltered. Further, we have chosen
\begin {eqnarray}\label{epsilon}
\varepsilon_{_{AA}}=\varepsilon_{_{BB}}=
2\varepsilon_{_{AB}}=\varepsilon.
\end {eqnarray}
This, alongside facilitating the phase separation, symmetrizes 
the model, thus providing an Ising like phase diagram. 
With regards to the density $\rho~(=\frac{N\sigma^3}{V}$, 
$N$ and $V$ being respectively the number of particles 
and volume of the system), we take a high value, viz., unity, 
so that interference with a vapor-liquid transition can be avoided.

\par
\hspace{0.2cm}The phase diagram using this model was studied 
before \cite{das3,das4}. In this work we revisit it to work 
out the system-size dependence of the critical temperature. 
Of course, due to the symmetry of the model, the critical 
concentration is automatically fixed at $x_{_A}^c=x_{_B}^c=1/2$ 
and a convenient definition of the order parameter then is 
\begin {eqnarray}\label{op}
m=\frac{1}{2}-x_{_A}.
\end {eqnarray}
Note that the concentration of species $\alpha$ is defined as 
$x_{_\alpha}={N_{_\alpha}}/{N}$, where $N_{_\alpha}$ is the 
number of particles of type $\alpha$ in the system. Like the 
earlier studies, here also we use Monte Carlo (MC) simulations 
in the semi-grandcanonical ensemble (SGMC) \cite{landau,frenkel} 
to understand phase behavior. In SGMC simulations, in addition 
to the standard particle displacement moves, one tries identity 
switches: $A \rightarrow B \rightarrow A$. After each trial, 
the move is accepted or rejected according to the standard 
Metropolis algorithm. For the latter moves, we have randomly 
chosen a particle and changed its identity. For this, one needs 
to incorporate the difference in chemical potential $(\Delta \mu)$ 
between the two species in the Boltzmann factor \cite{frenkel}. 
Due to the symmetry of our model, for $50:50$ composition above 
$T_{_c}$ and along the coexistence curve, we have $\Delta \mu=0$. 
One can also try random picking up of particles of a particular 
type and change its identity. In this case, however, the bias has 
to be taken care of by introducing an appropriate factor infront 
of the Boltzmann factor \cite{frenkel}.   

\par
\hspace{0.2cm}Due to the switch of identities in the SGMC 
simulations, one can obtain the probability distribution, 
$P(x_{_\alpha})$, for the concentrations, from the composition 
fluctuation. Above $T_{_c}$, $P(x_{_\alpha})$ will have a single 
peak structure and below $T_{_c}$, one expects a double peak form. 
The two peak structure of course implies phase coexistence with the 
locations of the peaks providing the coexistence composition as 
a function of temperature. Further, using normalized $P(x_{_A})$, one 
can calculate the concentration susceptibility as \cite{landau}
\begin {eqnarray}\label{chi1}
{k_{_B}}T\chi=N({\langle {x_{_A}}^2 \rangle}-
{\langle x_{_A} \rangle}^2).
\end {eqnarray}

\par
\hspace{0.2cm}Each run in the SGMC simulations started with 
a random $50:50$ mixture of $A$ and $B$ particles. In the 
displacement moves, randomly chosen particles were shifted in 
random Cartesian direction by a magnitude taken randomly from 
the range $\Big[0,\sigma/20\Big]$. All our simulations, MC as 
well as MD, are performed in cubic boxes of linear dimension 
$L$ with periodic boundary conditions in all directions and results 
are presented after averaging over multiple initial configurations. 

\par
\hspace{0.2cm}In case of dynamics, we have performed MD 
simulations \cite{frenkel,allen,rapaport} in the microcanonical 
(NVE) emsemble which was chosen for perfect preservation of 
hydrodynamics. Before the MD runs in the NVE emsemble, the 
system at that particular state point was equilibrated via Monte 
Carlo simulation in the canonical (NVT) ensemble and further 
thermalized via MD runs in the same ensemble by using Andersen 
thermostat \cite{frenkel}. In the MD simulations, Verlet velocity 
algorithm \cite{frenkel} was used to integrate the equations 
of motion. Based on the temperature, the time step, 
$\Delta t$, of integration was varied between $0.0025t_{_0}$ and 
$0.005t_{_0}$, where $t_{_0}=\sqrt{m_p\sigma^2/\varepsilon}$ is an 
LJ unit of time.

\begin{figure}[htb]
\centering
\includegraphics*[width=0.4\textwidth]{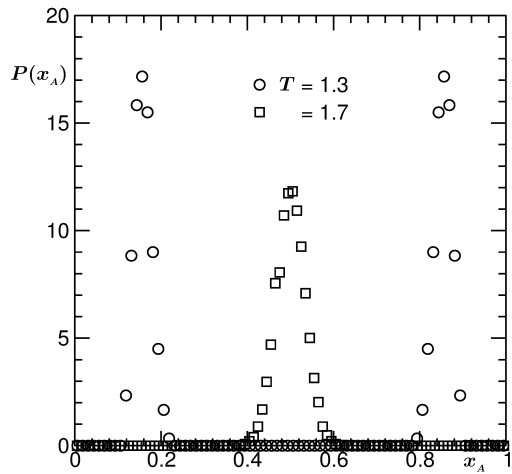}
\caption{\label{fig1} Plots of the concentration probability 
distribution, $P(x_{_A})$, vs $x_{_A}$, for two different 
temperatures. The system size considered was $L=10$.}
\end{figure}

\begin{figure}[htb]
\centering
\includegraphics*[width=0.4\textwidth]{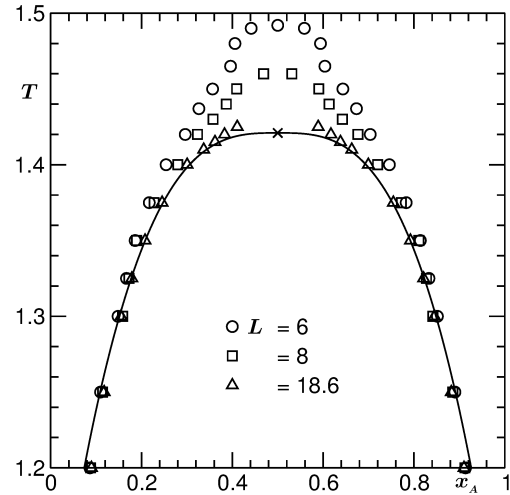}
\caption{\label{fig2} Coexistence curves for the binary LJ system 
in $T-x_{_A}$ plane, for three different system sizes. The 
continuous line is a fit to the largest system data, with the form 
$m \sim \epsilon^\beta$, by taking points in the region unaffected 
by finite size of the system and fixing $\beta$ to $0.325$. The 
cross corresponds to the coordinate $(x_{_A},T)\equiv (1/2,1.421)$, 
the thermodynamic critical point.}
\end{figure}

\par
\hspace{0.2cm}A standard practice in the computer simulations of 
dynamics is to calculate transport coefficient, $X$, using the 
Green-Kubo (GK) relation \cite{hansen}
\begin{eqnarray}\label{gk}
X(t)=\int_0^t dt' A(t'),~~ A(t')=\langle C(t')C(0)\rangle,
\end{eqnarray}
where $C$, e.g., for self-diffusion coefficient ($D$) is 
proportional to the individual particle velocity. Noting the 
definition of mutual diffusivity \cite{das3,das4}
\begin{eqnarray}\label{dab1}
D_{_{AB}}=\frac{{\mathscr L}}{\chi},
\end{eqnarray}
one identifies $\mathscr L$, the Onsager coefficient, as the 
pure dynamic quantity, the GK relation for which is \cite{hansen}
\begin{eqnarray}\label{ons1}
{\mathscr L}(t)=\left(\frac {t_{_0}\varepsilon}{k_{_B}NT\sigma^2}\right) 
\int_0^t dt' \langle {J_{_{AB}}^x}(t'){J_{_{AB}}^x}(0) \rangle,
\end{eqnarray}
where $\vec J_{_{AB}}$, the concentration current, is defined as 
\cite{hansen}
\begin{eqnarray}\label{ons2}
{\vec J}_{_{AB}}(t)={{x_{_B}}{\sum}_{i=1}^{N_{_A}}{\vec v_{_{i,A}}}(t)}
-{{x_{_A}}{\sum}_{i=1}^{N_{_B}}{\vec v_{_{i,B}}}(t)},~~
\end{eqnarray}
with ${\vec v}_{_{i,\alpha}}$ being the velocity of particle $i$ of 
species $\alpha$. Note that $J_{_{AB}}^x$ is the $x$-component of 
$\vec J_{_{AB}}$. The true values of these transport 
coefficients will be obtained in the limit $t\rightarrow \infty$. 
Instead of dealing with the velocities of all the particles, 
as seen in Eq. (\ref{ons2}), to facilitate faster calculation, 
one can use the momentum conservation to work with only one species. 
This will, in particular, be of advantage for very off-critical 
composition. This fact will be used in the derivation of the 
Einstein formula below.

\par
\hspace{0.2cm}Corresponding relations for viscosities can be 
written as \cite{hansen}
\begin{eqnarray}\label{shear1}
Y(t)={\left(\frac {t_{_0}^3 \varepsilon}{VT\sigma m_p^2}\right)
\int_0^t dt' 
\langle {P'_{ab}}(t'){P'_{ab}}(0)\rangle},
\end{eqnarray}
where $a,b \in [x,y,z]$ and $P'_{ab}$ are related to the elements 
$P_{ab}$, of stress tensor, having the expression
\begin{eqnarray}\label{shear2}
P_{ab}=\sum_{i=1}^N\Big[m_p{v_i^a}{v_i^b}+
{\frac {1}{2}} \sum_j^{'}{(a_{_i}-a_{_j})F_j^b}\Big].
\end{eqnarray}
Note that in Eq. (\ref{shear2}) $F_j^b$ is the $b-$component 
of the force acting on particle $j$ due to the others, $a_{_i}$ 
is the $a-$component of Cartesian coordinate of particle $i$ and 
similarly, $v_i^a$ is the $a-$component of the velocity of particle 
$i$. For $a=b$, one has $P=\langle P_{aa}\rangle$. In that case 
$P'_{aa}=P_{aa}-P$ and \cite{allen}
\begin{eqnarray}\label{shear3}
Y=\zeta+\frac{4}{3}\eta,
\end{eqnarray}
with $\eta$ being calculated from the correlator of the 
off-diagonal elements with $P'_{ab}=P_{ab}$. 

\begin{figure}[htb]
\centering
\includegraphics*[width=0.4\textwidth]{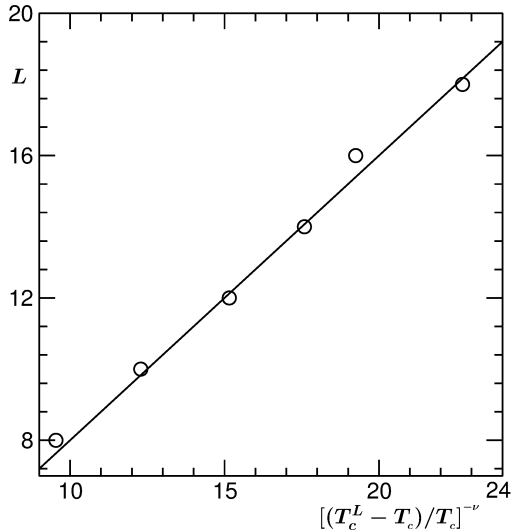}
\caption{\label{fig3} Plot of $L$ vs 
$[(T_c^L-T_{_c})/T_{_c}]^{-\nu}$. See text for the 
definition of $T_c^L$. The value of $\nu$ was fixed 
to $0.63$.}
\end{figure}

\begin{figure}[htb]
\centering
\includegraphics*[width=0.4\textwidth]{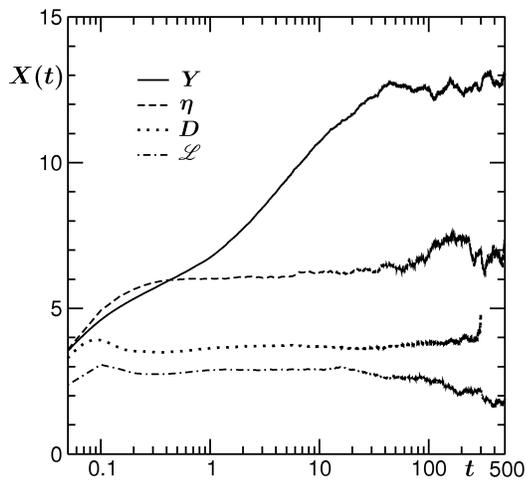}
\caption{\label{fig4} Plots of $Y(t)$, $\mathscr L(t)$, 
$\eta(t)$ and $D(t)$, for $L=10$. Results for $\mathscr L$, 
$\eta$ and $D$ were multiplied by suitable factors to make them 
visible on the range of ordinate. All results correspond 
to $x_{_A}=x_{_A}^c$ and $T=2.5$.}
\end{figure}

\par
\hspace{0.2cm}Because of the strong divergence that $\zeta$ 
exhibits, extremely long simulation runs are necessary, 
particularly in close vicinity of the critical point. For MD 
runs in microcanonical ensemble, which is needed for 
hydrodynamics, it is extremely difficult to keep the system 
temperature within acceptable fluctuation for long. Further, 
when the simulations are long, additional difficulty may appear 
in the accurate estimation of average pressure, due to 
truncation error, which is needed for the correlator in 
Eq. (\ref{shear1}). Essentially one requires long runs and 
averaging over huge number of independent initial configurations.

\par
\hspace{0.2cm}Instead of the GK relation, self-diffusivity 
is often calculated from the mean-squared-displacement (MSD), 
referred to as the Einstein relation, as \cite{hansen}
\begin{eqnarray}\label{ons3}
D=\lim_{t\rightarrow\infty} \frac{\langle|{\vec r}_{_i}(t)
-{\vec r}_{_i}(0)|^2\rangle}{6t},
\end{eqnarray}
which is a consequence of Fick's law of diffusion and provides 
information about random character of walk or displacement 
in the linear regime in long $t$ limit. As opposed to $D$, 
which is a single particle property, $D_{_{AB}}$ is related 
to the collective dynamics of the system. More precisely, 
it corresponds to the displacement of the centre of mass (CM) 
of one of the species which will be clear from its expression 
\cite{hansen,das5} analogous to Eq. (\ref{ons3}).

\par
\hspace{0.2cm}For a system with zero total linear momentum 
\begin{eqnarray}\label{ons4}
{\sum}_{i=1}^{N_{_A}}{\vec v_{_{i,A}}}(t)=-{\sum}_{i=1}^{N_{_B}}
{\vec v_{_{i,B}}}(t).
\end{eqnarray}
Defining the CM position $\vec R_{_\alpha}$ and CM velocity 
$\vec V_{_\alpha}$ of species $\alpha$ as 
\begin{eqnarray}\label{ons51}\nonumber
{\vec R_{_\alpha}}(t)=\frac{1}{N_{_\alpha}}{\sum}_{i=1}^
{N_{_\alpha}}{\vec r_{_{i,\alpha}}}(t),
\end{eqnarray}
\begin{eqnarray}\label{ons5}
~~{\vec V_{_\alpha}}(t)=\frac{1}{N_{_\alpha}}{\sum}_{i=1}^
{N_{_\alpha}}{\vec v_{_{i,\alpha}}}(t),
\end{eqnarray}
one writes the momentum conservation relation as
\begin{eqnarray}\label{ons6}
{N_{_A}}{\vec V_{_A}}(t)=-{N_{_B}}{\vec V_{_B}}(t).
\end{eqnarray}
From Eqs. (\ref{ons2}), (\ref{ons5}) and (\ref{ons6}), 
one obtains
\begin{eqnarray}\label{ons7}
{\vec J}_{_{AB}}(t)={N_{_A}}{\vec V_{_A}}(t),
\end{eqnarray}
which leads to 
\begin{eqnarray}\label{ons8}
{\mathscr L}(t)=\left(\frac{{t_{_0}}N_A^2\varepsilon}
{k_{_B}NT\sigma^2}\right)\int_0^t dt'
\langle {V_A^x}(t'){V_A^x}(0) \rangle.
\end{eqnarray}
Using 
\begin{eqnarray}\label{ons9}
\vec R_{_A}(t)-\vec R_{_A}(0)=\int_0^t dt'\vec V_{_A}(t')
\end{eqnarray}
in Eq. (\ref{ons8}), one gets \cite{hansen,das5}
\begin{eqnarray}\label{ons10}
{\mathscr L}(t)=\left(\frac{{t_{_0}}N_A^2\varepsilon}
{2k_{_B}tNT\sigma^2}\right)
\langle |R_{_A}^x(t)-R_{_A}^x(0)|^2\rangle.
\end{eqnarray}
A difference in the numerical factor ($2$ instead of $6$) 
between Eqs. (\ref{ons10}) and (\ref{ons3}) is due to the 
fact that here we are working with only one Cartesian component. 
In addition to the advantage of dealing with only one species 
in the system, Eq. (\ref{ons10}) provides data storage benefit, 
since one does not need to use data at infinitesimal intervals. 
This is useful for heavy computations like in critical dynamics. 

\par
\hspace{0.2cm}Similar expressions for viscosities can also be 
obtained \cite{hansen}, e.g., $\eta$ can be written as 
\begin{eqnarray}\label{shear4}
\eta(t)=\left(\frac {t_{_0}^3\varepsilon}{2k_{_B}tNT\sigma m_p^2}\right)
\langle |Q_{_{xy}}(t)-Q_{_{xy}}(0)|^2\rangle,
\end{eqnarray}
where the generalized displacement $Q_{xy}$ has the formula 
\cite{hansen}
\begin{eqnarray}\label{shear5}
Q_{_{xy}}(t)=\sum_{i=1}^N {x_{_i}(t)v_i^y(t)}.
\end{eqnarray}

\begin{figure}[htb]
\centering
\includegraphics*[width=0.4\textwidth]{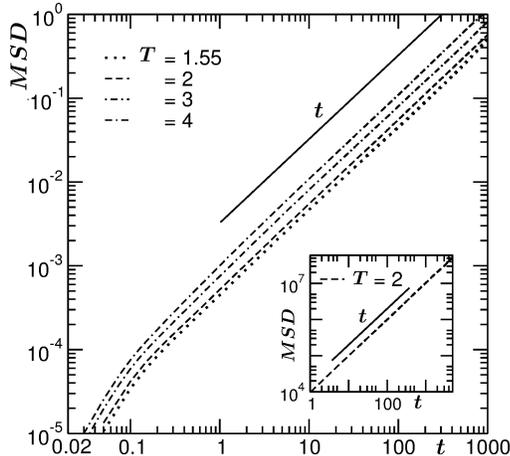}
\caption{\label{fig5}Mean squared displacements corresponding 
to $\mathscr L$ are plotted for different temperatures, at 
$x_{_A}=x_{_A}^c$, vs $t$. The inset shows corresponding result 
for $\eta$.}
\end{figure}

\par
\hspace{0.2cm}In order to relate the GK and Einstein relations, one 
needs to identify the relevant current, which, for self-diffusivity 
is the tagged particle velocity, for mutual diffusivity is the 
CM velocity of one of the species. The GK relation is the 
integration of the autocorrelation of this current. Corresponding 
displacement variable provides the Einstein relation. For 
shear viscosity the relevant quantity is the transverse current 
fluctuation \cite{hansen} $\frac{im_p}{k}\sum_{j=1}^Nv_ji^x(t)
exp[-i{\vec k}.{\vec r}_{_j}]$ ($k$ being the wave vector). 
The time derivative of this provides the elements of stress 
tensor and in the small $k$ limit, after imposing the total 
momentum conservation, one obtains Eq. (\ref{shear5}).

\par
\hspace{0.2cm}In this 
section we do not describe the methods of analysis of the 
simulation results. We will discuss them in the next section 
while presenting the results. Before proceeding to show the 
results, we set the values of $k_{_B}$, $m_p$, $\sigma$, 
$\varepsilon$ to unity and so of $t_{_0}$.

\section{Results}\label{results}
\par
\hspace{0.2cm}We start by showing plots of $P(x_{_A})$ vs $x_{_A}$ 
in Fig. \ref{fig1}, from SGMC simulations, for two different 
temperatures. It is clear from this figure that $T_{_c}$ lies 
in between these two temperatures. At very low $T$, it is possible 
that the system gets stuck in one of the phases because of high 
energy barrier coming from interfacial tension. In that case it 
is advisable to carry out biased simulations, e.g., successive 
umbrella sampling \cite{virnau}, to access the whole configuration 
space. On the other hand, one can use the model symmetry, if 
present, to identify the other phase. Indeed, in this figure we 
have used the fact 
\begin{eqnarray}\label{result1}
P(x_{_A})=P(1-x_{_A}),
\end{eqnarray}
because of which the data looks perfectly symmetric around 
$x_{_A}=1/2$.

\par
\hspace{0.2cm}In Fig. \ref{fig2} we show the coexistence curves 
in $T$ vs $x_{_A}$ plane for different system sizes. At very low 
temperature, there is a nice agreement of data from all the sizes. 
With the increase of temperature, one observes pronounced disagreement 
among various data sets, giving rise to strong system size 
dependence of the critical temperature. The value of $L-$dependent 
critical temperature \cite{Yethiraj}, $T_c^L$, is estimated to be the 
temperature at which there is a crossover from two-peak 
to single peak structure of $P(x_{_A})$. An initial guess for this 
can be facilitated by fitting the negative of the log of the low 
temperature distribution to Landau free energy form \cite{plischke} 
with symmetric double well having temperature dependent coefficients. 
Note that a true critical temperature is meaningful only in the 
limit $L \rightarrow \infty$. However, identification of $T_c^L$s 
will be helpful to obtain quantitative information on the critical 
singularities, in thermodynamic limit, from finite systems.

\par
\hspace{0.2cm}The continuous line in Fig. \ref{fig2} is obtained 
from the fit to the $L=18.6$ data in the finite-size unaffected 
region by fixing $\beta$ to $0.325$ and using $T_{_c}$ as an 
adjustable parameter. This provides $T_{_c}\simeq 1.421$. Note 
here that in an earlier work \cite{das3,das4}, from the crossing 
of Binder parameter \cite{binder} data from different values of 
$L$, $T_{_c}$ was obtained to be $1.423$. In that work no 
finite-size scaling analysis for Binder parameter crossing was 
done. However, in the same work, finite-size scaling analysis of the 
susceptibility data showed slightly better consistency with 
$T_{_c}=1.421$, when $T_{_c}$ was used as an adjustable parameter 
in data collapse experiment. So, we adopt this value for the rest 
of this paper.

\begin{figure}[htb]
\centering
\includegraphics*[width=0.4\textwidth]{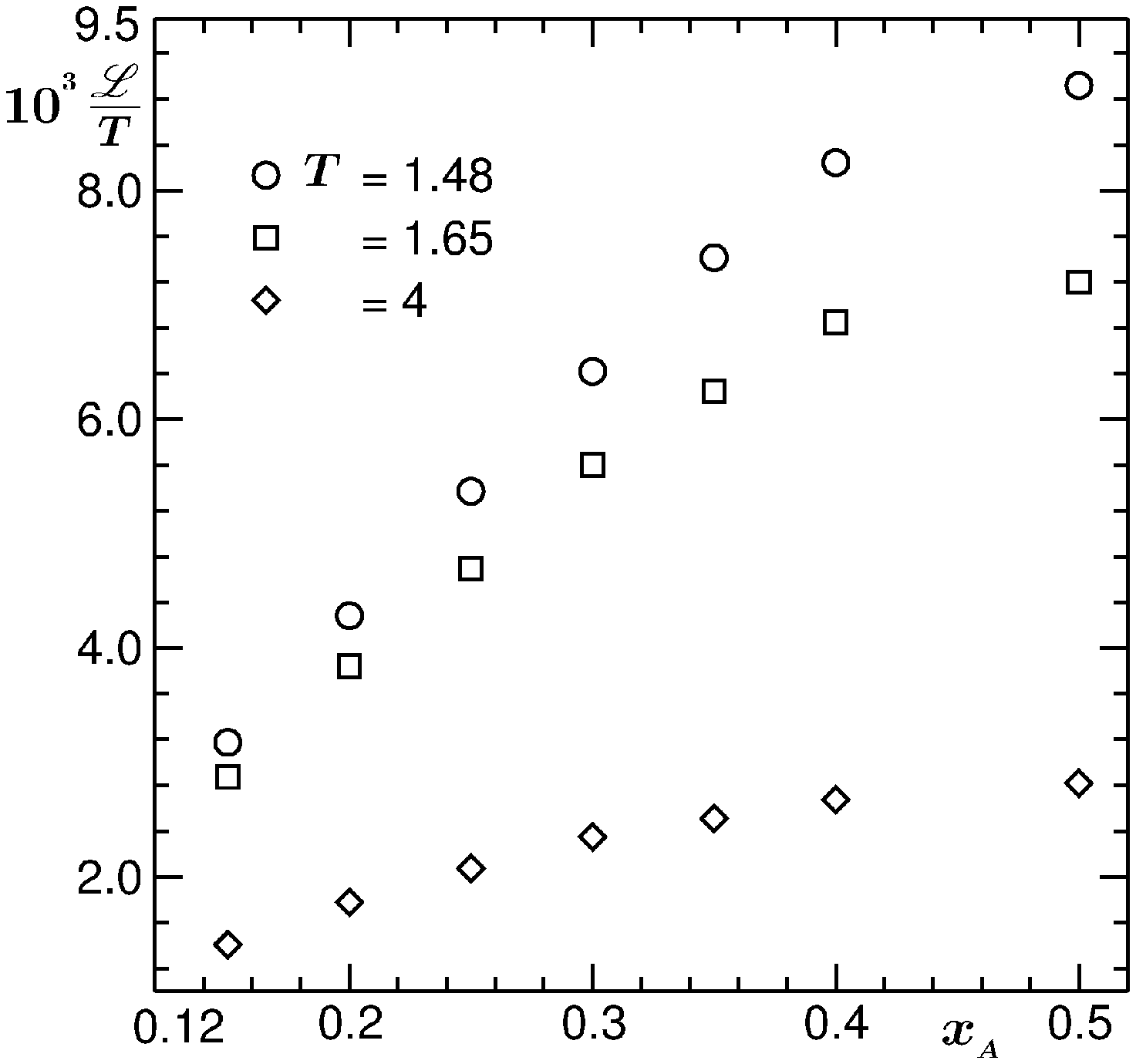}
\vskip 0.5cm
\includegraphics*[width=0.4\textwidth]{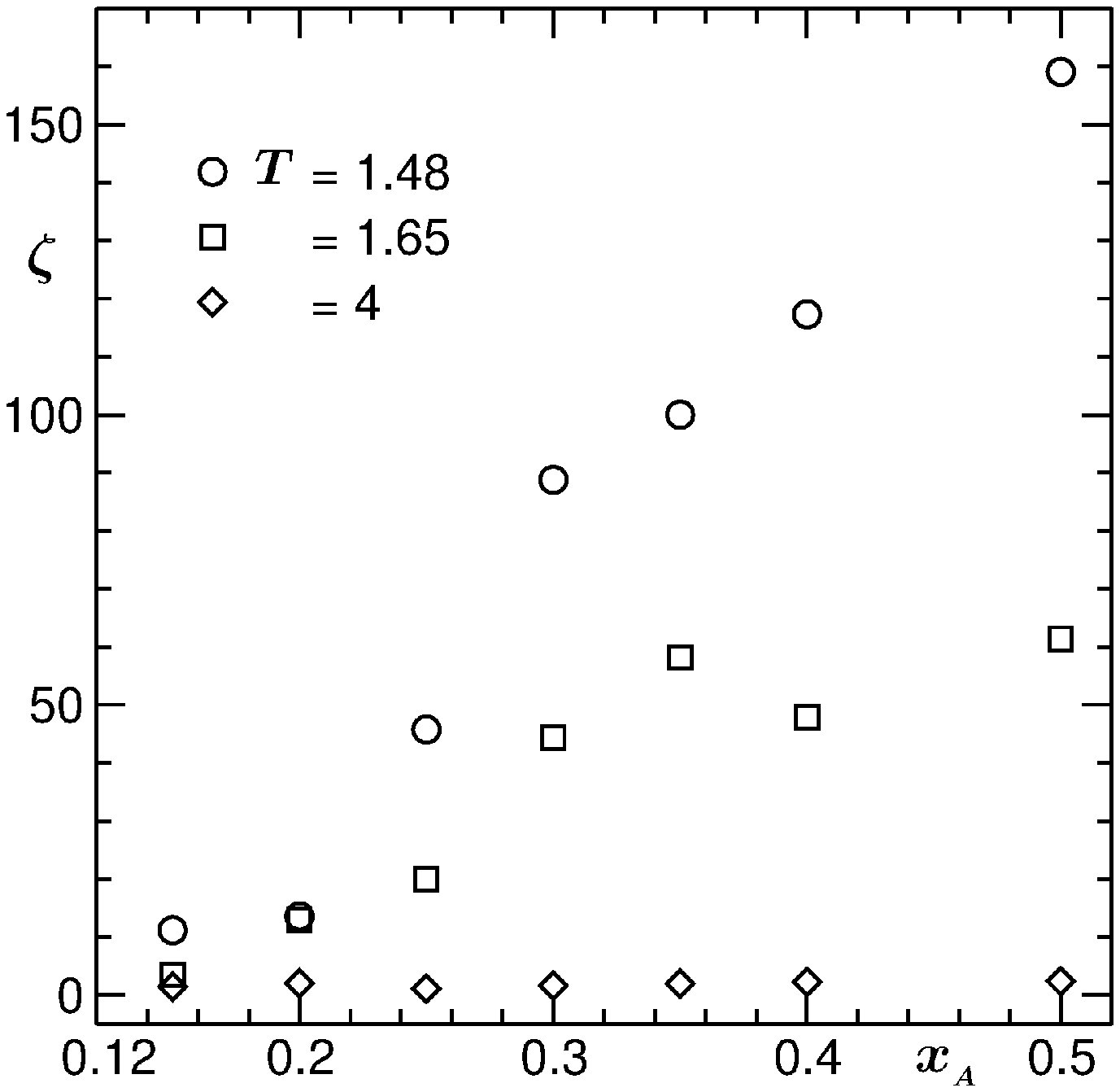}
\caption{\label{fig6}(a) Plots of ${\mathscr L}/{T}$ as a function 
of $x_{_A}$, for three different temperatures. All results are from 
same system size with $L=10$. (b) Same as (a), but for $\zeta$.}
\end{figure}

\par
\hspace{0.2cm}Considering the fact that at $T_c^L$, $\xi$ 
assumes the value $L$, one can write down
\begin{eqnarray}\label{result2}
L \sim \Big(\frac{T_c^L-T_{_c}}{T_{_c}}\Big)^{-\nu}.
\end{eqnarray}
In Fig. \ref{fig3} we have plotted $L$ vs $((T_c^L-T_c)/T_c)^{-\nu}$ 
by fixing $\nu$ to its Ising value and $T_{_c}$ to $1.421$. 
Consistency of the simulation data with the solid straight line 
confirms Eq. (\ref{result2}). We will use these values of $T_c^L$ 
later in the finite-size scaling analysis of transport coefficients. 
The random scatter seen in these data on both sides of the 
straight line is much smaller than the temperature fluctuation in 
NVE ensemble MD simulations.

\begin{figure}[htb]
\centering
\includegraphics*[width=0.4\textwidth]{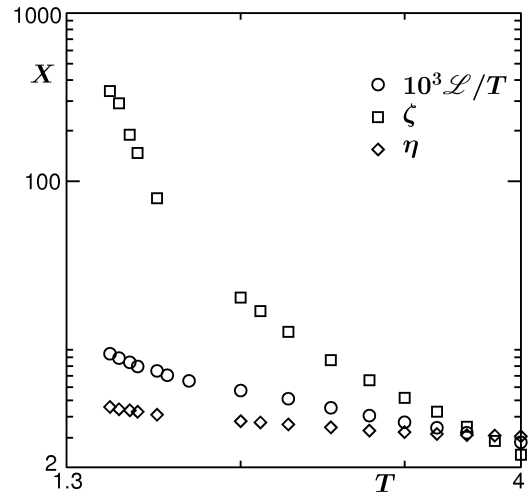}
\caption{\label{fig7}Plots of ${\mathscr L}/{T}$, $\eta$ and $\zeta$ 
as a function of $T$, at $x_{_A}=x_{_A}^c$. All results correspond to 
$L=10$.}
\end{figure}

\begin{figure}[htb]
\centering
\includegraphics*[width=0.4\textwidth]{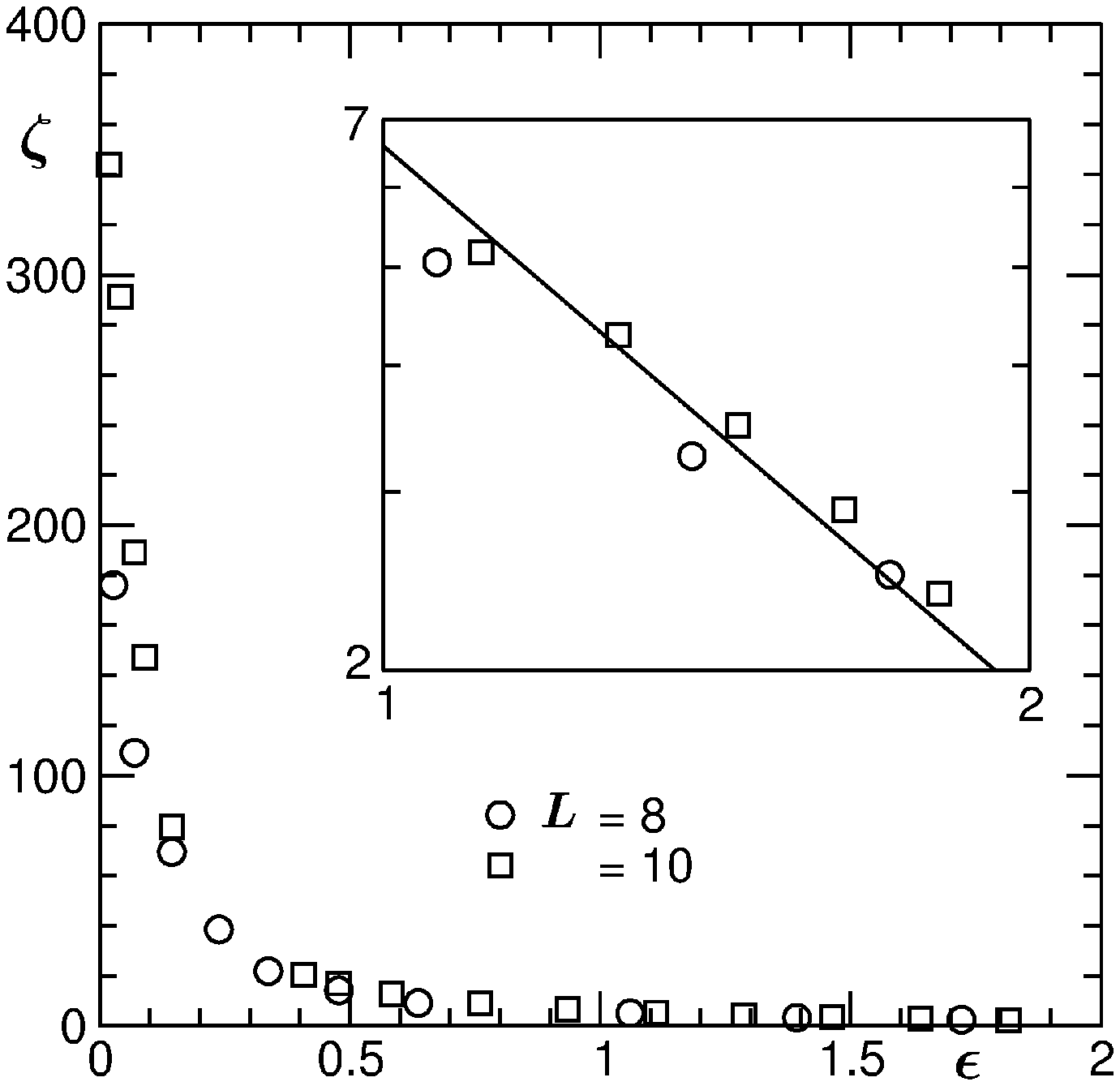}
\vskip 0.5cm
\includegraphics*[width=0.4\textwidth]{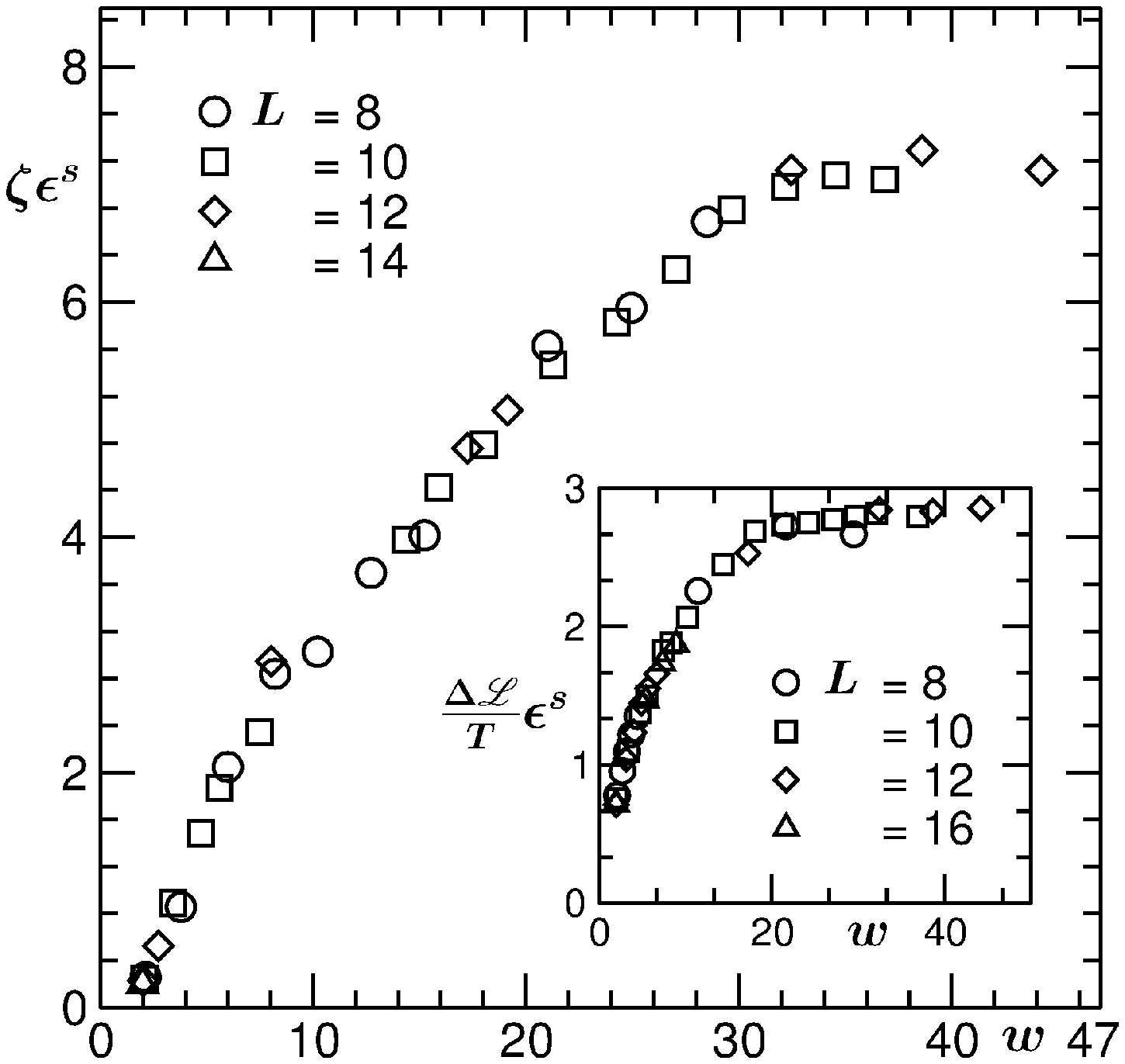}
\caption{\label{fig8}(a) Plots of $\zeta$ vs $\epsilon$, along 
$x_{_A}=x_{_A}^c$, for two different system sizes. The solid 
line in the inset has a power-law divergence with exponent $1.82$. 
(b) Finite-size scaling plot for $\zeta$, viz., $\zeta\epsilon^s$ 
vs $w$, with $s=\nu x_{_\zeta}=1.82$. Data from four different $L$ 
values are used. Inset: Same as the main figure, but for $\mathscr L$. 
In this case $s$ has a value $0.567$.}
\end{figure}

\par
\hspace{0.2cm}On dynamics, we start by demonstrating how various 
quantities were obtained. In Fig. \ref{fig4} we show the 
integrations of the autocorrelations for $\mathscr L$, $\eta$, 
$Y$, and $D$ as functions of time. These quantities were 
estimated from the pleatue region. All these results correspond to 
$x_{_A}=1/2$, $T=2.5$ and $L=10$. The relaxation of the correlators for 
different quantities at different times is indicative of varying 
critical enhancement. The result for $D$ appears smoother than 
the others. This is due to the difference between collective 
system property and individual particle property. In the latter 
case, there is scope of self averaging over all the particles 
in the system. The lack of this for collective properties 
make the estimation of these quantities significantly difficult, 
particularly in the close vicinity of critical point where relaxation 
time diverges.

\par
\hspace{0.2cm}In Fig. \ref{fig5} we show the MSDs for 
$\mathscr L$ at different temperatures along $x_{_A}^c$. The 
corresponding result for $\eta$ is shown in the inset. At late 
time nice linear behavior is seen, implying Einsteinian diffusion 
of relevant variables. From the time derivative of these 
results, in long time limit, the transport coefficients can be 
estimated which, we have checked, are in good agreement with 
the estimates from Fig. \ref{fig4}.

\begin{figure}[htb]
\centering
\includegraphics*[width=0.4\textwidth]{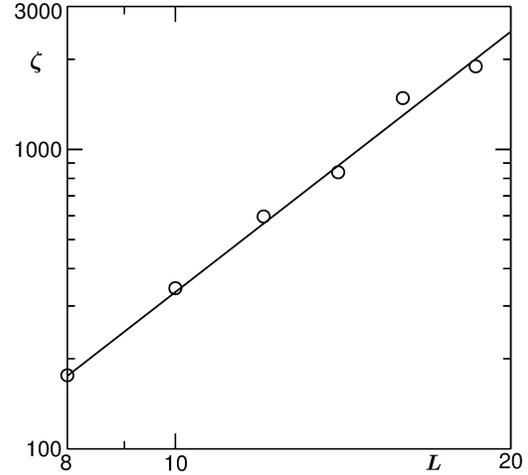}
\vskip 0.5cm
\includegraphics*[width=0.4\textwidth]{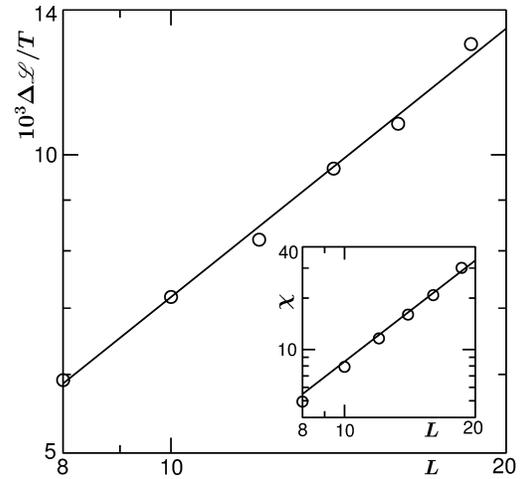}
\caption{\label{fig9}(a) Plot of $\zeta$, calculated at $T_c^L$s, 
as a function of $L$. The solid line represents power-law with 
exponent $2.89$. (b) Same as (a), but for ${\Delta {\mathscr L}}/{T}$. 
Here the solid line has exponent $0.9$. The inset shows same 
exercise for $\chi$.}
\end{figure}

\begin{figure}[htb]
\centering
\includegraphics*[width=0.4\textwidth]{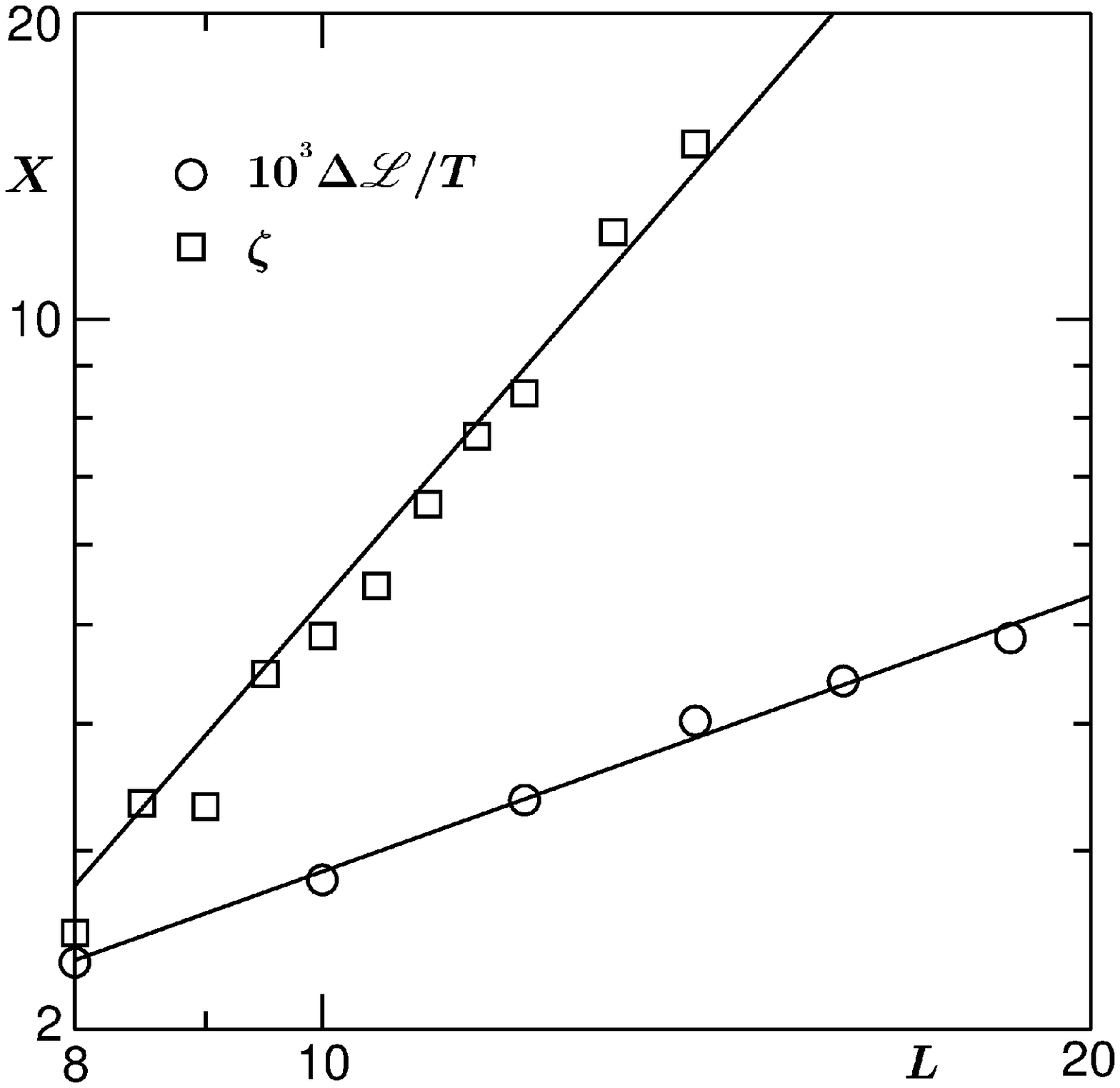}
\caption{\label{fig10} Same as Fig. \ref{fig10}, but the 
quantities were calculated at $T_c^L(f)$ with $f=65$.}
\end{figure}

\par
\hspace{0.2cm}In Fig. \ref{fig6}(a) we show the plots of 
${\mathscr L}/{T}$ as a function of $x_{_A}$, for three different 
temperatures. Note that we have confined ourselves to $x_{_A}\le0.5$. 
The result will be mirror image on the other side because of the 
symmetry of the model. The reason for choosing ${\mathscr L}/{T}$, 
instead of only $\mathscr L$, is clear from Eqs. (\ref{intro2}), 
(\ref{exponent4}) and (\ref{dab1}) which suggest
\begin{eqnarray}\label{result3}
\frac{\mathscr L}{T}=Q\epsilon^{-\nu_{_\lambda}};
~\nu_{_\lambda}=0.567,
\end{eqnarray}
where $Q$ is a critical amplitude. Enhancement of 
${\mathscr L}/{T}$ with the approach towards the critical 
composition is clear which is more and more prominent for 
temperatures closer to the critical value. Interestingly, 
this enhancement even for $T=4$ is significant and perhaps 
signals a very wide critical range. In Fig. \ref{fig6}(b) 
we present analogous results for $\zeta$. Qualitative behavior 
of the data here is same as Fig. \ref{fig6}(a), but rise in 
$\zeta$ as a function of $x_{_A}$ as well as $T$ appears 
much stronger than $\mathscr L$, implying a larger value of 
critical exponent for $\zeta$. This is, of course, the 
prediction of theories which we will quantify in this work. 
Note that an apparently flat look of data for $T=4$ in this 
case is due to the large scale of the ordinate.

\par
\hspace{0.2cm}A comparative picture of the critical enhancement 
of various transport coefficients, along the critical 
concentration, is shown in Fig. \ref{fig7}. Here, in addition to 
$\zeta$ and $\mathscr L$, we have included $\eta$ also. Data for 
$\mathscr L$ has been multiplied by $1000$. It is 
clearly seen that the enhancement of $\zeta$ is much stronger than 
$\mathscr L$. On the other hand, data for $\eta$ appears rather 
insensitive to the approach to criticality. This is consistent 
with the prediction of a very small value for the corresponding 
critical exponent. Considering this and combining with the 
fact that temperature fluctuation in NVE ensemble is rather 
high, we do not aim to proceed further with detailed analysis 
for the critical behavior of $\eta$. For $D$, on the other 
hand, there is no theoretical expectation for critical enhancement 
which we will demonstrate towards the end of the paper.

\begin{figure}[htb]
\centering
\includegraphics*[width=0.4\textwidth]{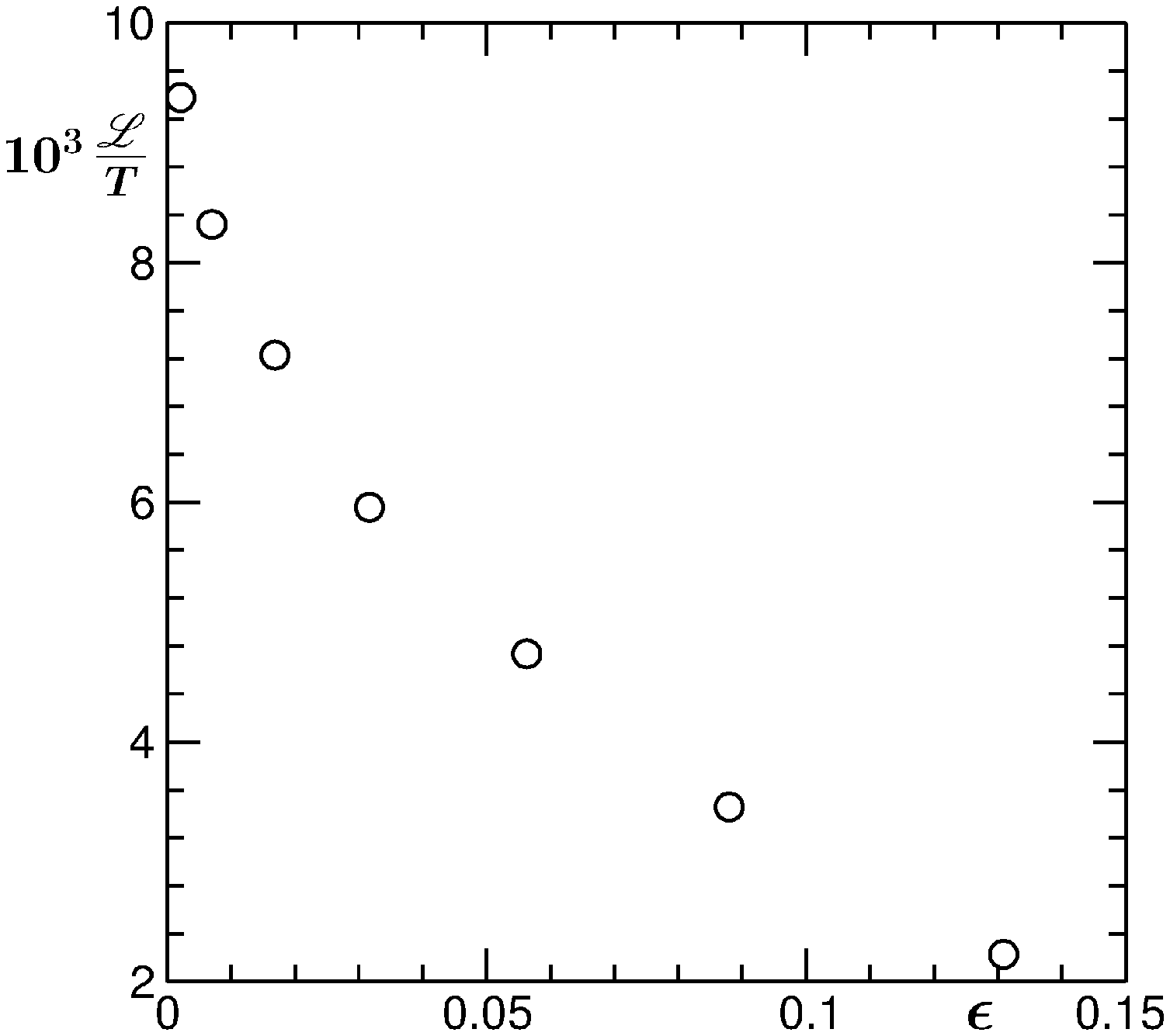}
\caption{\label{fig11}Plot of ${\mathscr L}/{T}$ as a function 
of $\epsilon$, for $T<T_{_c}$, along the B-rich branch of the 
coexistence curve. The system size was chosen to be $L=10$.}
\end{figure}

\par
\hspace{0.2cm}In Fig. \ref{fig8}(a) we present results for $\zeta$ 
vs $\epsilon$, along $x_{_A}=x_{_A}^c$, for two different system 
sizes. The discrepancy between the two data sets for smaller values 
of $\epsilon$ is due to finite-size effects which appear rather 
strong that can be appreciated from the fact that results from 
the two systems agree only at very large value of $\epsilon$. In 
the latter region, shown on double-log scale in the inset, the 
data are consistent with the solid line which has a power-law with 
theoretical exponent $\nu x_{_\zeta}=1.82$. To confirm this critical 
exponent further, we take help of the following finite-size scaling 
method.

\begin{figure}[htb]
\centering
\includegraphics*[width=0.4\textwidth]{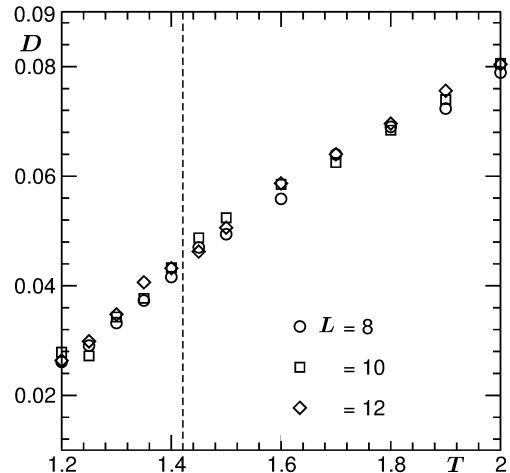}
\caption{\label{fig12}Plot of self-diffusivity $D$, for species 
$A$, as a function of $T$. Results for few different system sizes 
are shown. For $T<T_{_c}$, we have chosen the B-rich branch of the 
coexistence curve. The vertical dashed line represents the value 
of $T_{_c}$.}
\end{figure}

\par
\hspace{0.2cm}Since fluctuation can never become truly critical in a 
finite system, the enhancement 
\begin{eqnarray}\label{fss1}
X \sim \epsilon^{-s}
\end{eqnarray}
gets suppressed there. Nevertheless, it is possible to obtain 
information about the exponent $s$ and other relevant quantities 
in critical phenomena via appropriate scaling analysis 
\cite{landau,fisher2} involving finite values of $L$. For 
$L<\infty$, one introduces a finite-size scaling function $Z$ 
to write 
\begin{eqnarray}\label{fss2}
X=Z(w)\epsilon^{-s},
\end{eqnarray}
where $Z$ is independent of system size and depends upon the 
dimensionless variable $w=L/\xi$, $\xi$ having the thermodynamic 
limit behavior. Thus, in a plot of $X\epsilon^s$ vs $w$, data for 
different values of $L$ will collapse onto a master curve, of 
course, if the exponent $s$ is appropriately chosen. For far away 
from $T_{_c}$, where $\xi << L$, one does not expect finite-size 
effects. In that limit, $Z(w)$ will be a constant equal to the 
critical amplitude of the quantity under study. For $T$ very close 
to $T_{_c}$, finite-size effects will certainly become important, 
thus $X$ will tend towards a constant, allowing one to write
\begin{eqnarray}\label{fss3}
Z(w)\sim w^{s/\nu}\sim \epsilon^s.
\end{eqnarray}

\par
\hspace{0.2cm}We do an exercise along the line described above, 
in the main frame of Fig. \ref{fig8}(b), for $\zeta$. Here the 
value of $s$ was set to $1.82$ which provides excellent collapse 
of data from many different system sizes. Flat portion, in this 
figure, in the large $w$ limit, provides the value of critical 
amplitude to be $\simeq 7.07$.

\par
\hspace{0.2cm}Similar analysis for $\mathscr L$ was done 
previously \cite{das3,das4}. For the sake of completeness, we 
show it in the inset of Fig. \ref{fig8}(b), this time with more 
data points. Previously, strong background in $\mathscr L$, 
coming from small length fluctuations, was observed \cite{das3,das4}, 
so that in the critical vicinity one should write
\begin{eqnarray}\label{fss4}
{\mathscr L}={\mathscr L}_b+\Delta{\mathscr L},
\end{eqnarray}
where ${\mathscr L}_b$ and $\Delta{\mathscr L}$ are respectively 
the background and critical contributions. To obtain information 
about the critical exponent, in such cases, one needs to subtract 
the value of ${\mathscr L}_b$ appropriately. This was done by using 
${\mathscr L}_b$ as an adjustable parameter in the finite-size 
data collapse experiment by fixing $x_{_\lambda}$ to $0.567$. 
The value that provides best collapse is \cite{das3,das4} 
${\mathscr L}_b\simeq 3.3\times 10^{-3}$, giving 
$Q\simeq 2.8\times 10^{-3}$ which can be appreciated from the 
plot presented in the inset. In the following we will use this 
number for ${\mathscr L}_b$ and check via other type of analysis 
if the exponent value $x_{_\lambda}=0.567$ is correct which was 
fixed in this case.

\par
\hspace{0.2cm}Next, we exploit the scaling behavior of $T_c^L$ with 
$L$, demonstrated in Fig. \ref{fig3}, to obtain further 
confirmation on the critical divergences of these quantities. 
By observing Eq. (\ref{result2}), while dealing with data 
calculated at $T_c^L$s, one can write down
\begin{eqnarray}\label{fss5}
X \sim L^{-s/\nu}.
\end{eqnarray}
In Fig. \ref{fig9}(a) we plot $\zeta$ vs $L$ on a double-log 
scale using data at $T_c^L$s. The consistency of the simulation 
data with the solid line, bearing the theoretical exponent 
$2.89$, is clearly visible. Here note that for a binary fluid 
the value of the exponent $x_{_\zeta}$ was pointed out 
\cite{jkb1,jkb2} to be closer to $3$. This difference, however, 
is difficult to confirm from the quality of our simulation data. 
In Fig. \ref{fig9}(b), we show ${\Delta \mathscr L}/{T}$ vs $L$, 
again on double-log scale. Here also the data are consistent with the 
theory, represented by the solid line. This, in addition, confirms 
the choice of ${\mathscr L}_b$, quoted earlier. In the inset of 
Fig. \ref{fig9}(b) we do same exercise for $\chi$ demonstrating 
consistency with theory, thus confirming that $x_{_D}=1.07$.

\par
\hspace{0.2cm}Starting from $T_c^L$, one can define effective 
critical temperature as \cite{das1,roy1}
\begin{eqnarray}\label{fss6}
T_c^L(f)=T_{_c}+f(T_c^L-T_{_c}),
\end{eqnarray}
which, for $f=0$, is the thermodynamic critical point and 
for $f=1$, is just $T_c^L$. Scaling of $T_c^L(f)-T_{_c}$, for any 
arbitrary value of $f$, with $L$, should be same as 
Eq. (\ref{result2}). Usefulness of such effective critical temperature 
can be understood as follows. For significantly large values of $L$, 
if it does not become possible to calculate a quantity at $T_c^L$, 
which indeed is true for transport properties with strong critical 
enhancements, one can choose a suitably large value of $f$ to avoid 
computational effort, of course, if the critical range is extended 
that far. In this paper, we will use a rather large value of $f$, 
to obtain information on the critical range. 

\par
\hspace{0.2cm}In Fig. \ref{fig10} we show $\zeta$ as well as 
${\Delta \mathscr L}/{T}$ as functions of $L$, with quantities 
calculated at $T_c^L(65)$. The linear looks of both the data sets 
on a log-log scale are indicative of power-law behaviors. The 
solid lines there have corresponding theoretical exponents. The 
temperatures included in this figure, lie in the range $[1.91,3.87]$. 
Nevertheless, the consistency of the simulation results with theory 
confirm that the critical range for this model is rather large, 
a hint of this was already obtained from Fig. \ref{fig8}. For a 
binary fluid it is expected that the critical range will be wider 
than a vapor-liquid transition. However, the range we see 
in this work is very high which could possibly be attributed to 
the symmetry of the model. 

\par
\hspace{0.2cm}In Fig. \ref{fig11} we show ${\mathscr L}/{T}$ as a 
function of $\epsilon$, for temperatures below $T_{_c}$, for $L=10$, 
calculated at state points along the B-rich branch of the 
coexistence curve. Enhancement of the quantity with the increase 
of temperature is clearly visible. Due to lack of data from different 
system sizes we do not aim to quantify the exponent in this case. 
Note that the background contribution below $T_{_c}$ need not be 
the same as above. Also, while doing finite-size scaling exercise, 
one needs to consider the finite-size effects in the coexistence 
curve as well, as seen in Fig. \ref{fig2}. All these put together, 
it is a substantial task which we leave out for a future work. 
Nevertheless, the information provided in Fig. \ref{fig11} 
(as well as in Fig. \ref{fig12}) is expected to be important 
in other studies. Noting the simplicity as well as reasonably 
realistic nature, the model is useful in the understanding of 
hydrodynamic effects in kinetics of phase separation. For 
quantitative understanding of the latter, one requires information 
about various transport coefficients for coexistence composition.

\par
\hspace{0.2cm}Finally, in Fig. \ref{fig12} we present the 
self-diffusion coefficient, $D$, for species $A$, at temperatures 
above and below $T_{_c}$. We have included results from few 
different system sizes. No systematic system size dependence is 
visible. This, combined with the overall nature of data, do not 
suggest any critical singularity for this quantity. 

\section{Conclusion}\label{conclusion}
\par
\hspace{0.2cm}In summary, we have studied phase behavior and 
dynamics in a symmetric binary Lennard-Jones model exhibiting 
liquid-liquid transition. The phase diagram was studied via 
Monte Carlo simulations in a semi-grandcanonical ensemble 
\cite{landau} whereas we have used molecular dynamics 
simulations \cite{frenkel} in the microcanonical ensemble to 
obtain information on dynamics. 

\par
\hspace{0.2cm}Various transport properties, viz., self and 
mutual diffusivities, shear and bulk viscosities were 
calculated using Green-Kubo as well as Einstein relations 
\cite{hansen}. We have explored a wide range of 
state points on and around the coexistence curve. Particular 
focus was on the understanding of critical singularities. 

\par
\hspace{0.2cm}It is observed that the self diffusivity is 
insensitive to the critical point and shear viscosity exhibits 
only weak critical enhancement. The critical behaviors of 
mutual diffusivity as well as bulk viscosity were quantified 
via various finite-size scaling methods. Our results are nicely 
in agreement with the predictions of mode coupling and 
dynamic renormalization group theories \cite{hohenberg,ferrell,
olchowy,onuki1,anisimov,kadanoff,mistura,folk,onuki2,hao,jkb1,jkb2}. 

\par
\hspace{0.2cm}Interestingly, critical range for these 
transport properties appear to be very large. It is not clear 
to us whether this is due to the symmetry of the model or the particular 
interatomic potential. This finding 
is consistent with one \cite{meier} of the previous studies for a 
vapor-liquid transition, with Lennard-Jones interaction, where 
there is no such symmetry. 
\par
\hspace{0.2cm}
For bulk viscosity it needs to be seen whether the Lennard-Jones 
interaction allows for no other background term until the critical 
enhancement has become of the same order as the shear viscosity. In 
Ref. [\cite{meier}], even though strong enhancement far away form 
$T_c$ was noticed, no quantitative study with respect to the 
critical divergence was done. Recently we \cite{jia} have undertaken 
independent study of bulk viscosity for vapor-liquid transition, 
again with Lennard-Jones potential. Quantitative outcome from there can possibly 
provide some hint along this line.

\par
\hspace{0.2cm}In future, we will take up the issue of quantifying 
the exponents below $T_{_c}$. In that case, it will be interesting 
to examine the existence of universal amplitude ratios, as is true 
in static critical phenomena \cite{privman}. To establish this, 
one of course needs to study various different models. 

\vskip 1.0cm

\section*{Acknowledgement}\label{acknowledgement} 
SKD and SR acknowledge financial support from the Department of 
Science and Technology, India, via Grant No SR/S2/RJN-$13/2009$. 
SR is grateful to the Council of Scientific and Industrial 
Research, India, for their research fellowship.
\vskip 0.5cm
\par
$*$~das@jncasr.ac.in

\end{document}